Title: Extended Jahn-Teller effects in fermion-boson systems revisited
Author: Mladen Georgiev (Institute of Solid State Physics, Bulgarian Academy of Sciences,
   1784 Sofia, Bulgaria)
Comments: 6 pages including 3 figures, all pdf format
Subj-class: physics

Following earlier suggestions, we now redefine our proposal for extending the Jahn-Teller coupling beyond the framework of the electron-phonon interactions so as to cover fermion-boson interactions on a broader energy scale. Our basic unit is an extended "vibronic polaron" (VP), a fermion, itinerant or localized, along with its associated distortion of its boson cloud. A fingerprint of the itinerant unit is the helical form of VP propagation, else the occurrence of an orbital angular momentum for the free carrier. Examples for extended VP's are provided by LG-mode laser beams, and possibly neutrino-meson coupled units. The VP theory extends to supersymmetry where it accounts for symmetry breakings, both external and internal ones.

1. Apologia

Recently we proposed that the helical form of light propagation may result from symmetry breaking similar to the Jahn-Teller (JT) effects [1]. In a broader sense, the JT effects may be extended so as to comprise the mixing of fermion states through their coupling to a boson field, not necessarily a phonon field. In the electron-phonon version, the entity formed through phonon mixing of electronic bands is called a JT polaron.

We apologize for revisiting the principal topic again albeit from a different perspective: We had shown earlier that the driving force for the occurrence of an orbital angular momentum of a free vibronic polaron (VP) was the off-center displacement of the fermion charge due to boson coupling. In addition, the off-centered species was known to perform an orbital-like reorientational motion around the central site. Ultimately, this resulted in helical propagation of the fermion-boson coupled species. Among fermion-boson coupled components to be subject of a JT study we name but a few: fermions (electrons, neutrinos, nucleons), bosons ($\pi$-mesons, phonons, photons), etc.

We also raise the question as to the form of boson features in the parallel event. There is one comprehensive study using the Variational Ansatz [2]. The method allows for the definition of an average phonon coordinate $Q_{\alpha\beta}$ which has been computed numerically for antiadiabatic (AA) and semibound (SB) vibronic polarons in 1D. Numerical calculations reveal coordinate maximums at the center of the Brillouin zone (AA) and at $k = \pm\frac{1}{2}\pi$ away from the center (SB), as in harmonic oscillator theory [3]. They reveal extremal configurational structures.

We have also observed a link with supersymmetry (susy): Suppose an elementary band-mixing step that involves a fermion in an upper band $\beta$ which on emission of a phonon $\gamma$ is brought down to a lower band $\alpha$. What is essential is that two particles are born in the process, a $\gamma$ phonon and an $\alpha$ fermion, both at the expense of a $\beta$ fermion. Apparently, the $(\alpha,\gamma)$ fermion-boson pair is composed of two supersymmetric partners. In as much as the whole JT process leads to the breaking of an external symmetry, the explicit reference to super-symmetry implies a link to an internal symmetry to be broken by JT as well [4].

In revisiting the helical propagation problem, we are motivated by the increased interest in fundamental optics, on the one hand, and by the recent interest in using neutrino beams for extraterrestrial communications, on the other [5]. Not least important is our obtained result that helical propagation is perhaps an essential feature of coupled fermion-boson propagation which is, thereby, a fingerprint of supersymmetry [1].

2. Vibronic polarons

The vibronic polaron (VP) is a broken-symmetry entity composed of an electron coupled to its associated lattice deformation produced as a pair of degenerate or nearly-degenerate electronic states are mixed by appropriate phonon mode(s) giving rise to Jahn-Teller (JT) or Pseudo-Jahn-Teller (PJT) distortions, respectively [6]. Generalizations to more components are readily conceivable.

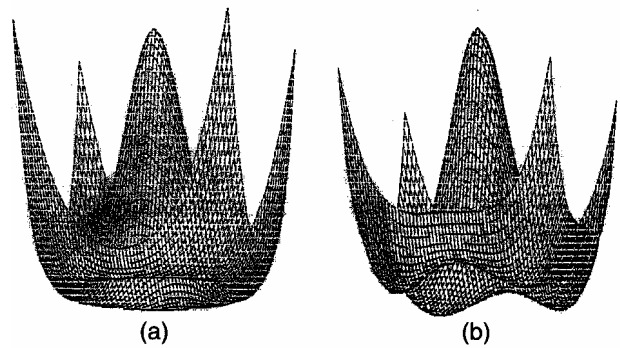

Figure 1

Sombrero type vibronic potential up to first order (a) and up to third order (b) in the electron-phonon mixing constants. The brim is smooth in (a) and embroidered by rotational barriers in (b). (Courtesy reference [6]).

There are itinerant VP units (mostly antiadiabatic ones) or localized VP units (often adiabatic ones). In view of the internal symmetry of JT or PJT vibronic polarons, an extension may prove feasible in which the electrons are upgraded to fermions while the phonons are upgraded to bosons. The symmetry lowering is the result of the vibronic mixing interactions (JT or PJT). A particular case is the off-center displacement which occurs as a result of mixing PJT interactions at a central lattice site driving the occupying ion off site [7]. The phonon version will be upgraded to a fermion-boson off-center displacement. The quantum-mechanical off-center effect will be seen to be at the center of our present discussion. Its Hamiltonian is given by equation (1) to appear below, while its eigenfunction is [1]:

$\psi(z,\varphi) = A\Phi(\varphi) \exp(ikz)$

The φ-angular part has been dealt with in equatorial plane. Its eigenfunctions are Mathieu's periodic functions, while its eigenvalues fall within Mathieu's rotational energy bands [7]. The z-dependent part of the Hamiltonian simplifies to:

$$H_{vib,z} \sim -(\hbar^2/2I)(\partial^2/\partial z^2)$$

where I is the rotational momentum of the off-center entity. Figure 1 illustrates the angular vibronic potentials in the off-center volume of a 2D vibronic polaron.

## 3. Helical light beams

There have been early suggestions and interest in the conceivable helical form of light propagation [8]. To that extent Pointing realized at the turn of the 20th century that circularly polarized light has (spin) angular momentum. Later studies established that light beams with an azimuthal phase dependence $\exp(-i\ell\phi)$ exhibit an orbital angular momentum independent of the polarization state. As with the circularly polarized light, the sign of the orbital angular momentum indicates a handedness of the beam with respect to its direction of propagation. These observations have been relatively untypical, though the first systematic experiments dating back to the early '90s recognized light beams that carried angular orbital momentums independent of polarization [9]. For helical phase fronts, the Pointing vector has an azimuthal component that produces an orbital angular momentum parallel to the propagation axis which momentum circulates about the beam axis; these beams are said to contain an optical vortex.

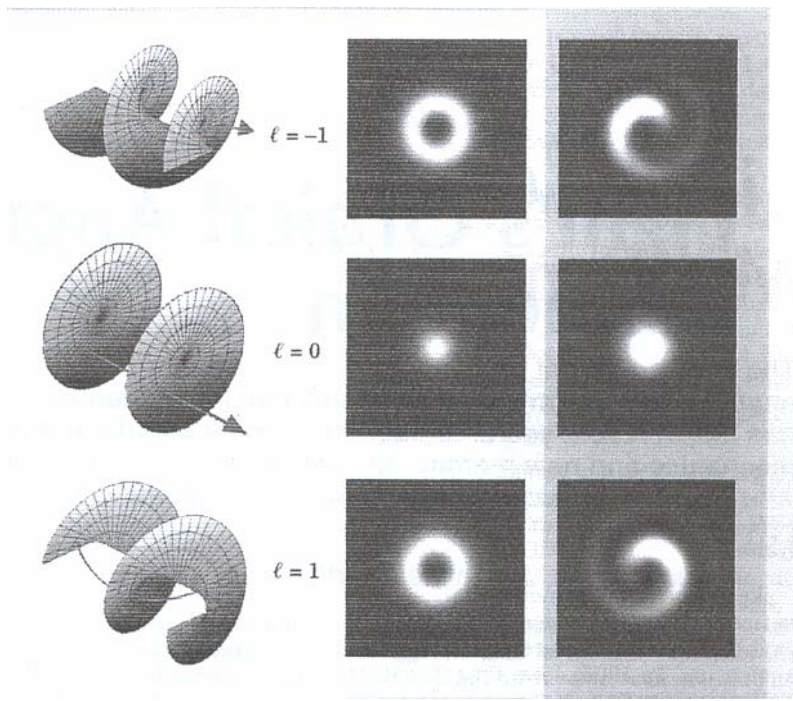

Figure 2

Orbital angular momentum of a light beam independent of the beam polarization. It arises from helical phase fronts (left panel). Beams with helical wave fronts have annular intensity profiles (central panel). When a helical beam is made to interfere with a plane wave, it results in a spiral intensity profile (right panel). (After reference [9]).

The most common form of a helically phased beam is the LG laser mode. This mode having an explicit exp(-i$l\phi$) phase factor, it is the natural choice for generating and studying light beams with an orbital angular momentum. From this rather scarce introduction which can be found extended elsewhere [9], it follows that the helical phase form depends on the generation conditions rather than on any inherent property of a light beam. The spin angular momentum depends only on the polarization of a light beam, not on its phase. The orbital angular momentum depends on the phase factor as well. Spin and orbital angular momentum are similar in many ways, though they have different interaction properties. Nevertheless, there is a substantial difference too: while the spin angular momentum has only two independent states corresponding to left- and right-handed circular polarization, the orbital angular momentum has an unlimited number of states corresponding to all integer $l$. The helical light forms are exemplified in Figure 2 for L = 1 ($l$ = - 1, 0, + 1) [9].

Generally, the orbital angular momentum arises whenever the phase fronts of a light beam are not perpendicular to the direction of propagation. In its turn, our approach points to the quantum mechanical off-center effect as the generator of light orbital chirality in the orbital angular momentum. One may conclude that the coupling of the laser field in the LG mode (Laguerre-Gaussian mode) results in the off-center displacements of fermions in the active medium giving rise to phase fronts off-perpendicular to the direction of propagation. One should follow the standard procedures to calculate a fermion-boson coupling constant (mixing constant for electronic states) in order to assess the magnitude of the JT effects and thereby the liability of the off-center hypothesis.

## 4. Supersymmetric partners

We start by reminding of the cooperative Jahn-Teller effect Hamiltonian [10,11]:

$$H = H_e + H_{ph} + H_{int} \quad (1)$$

where in site and momentum representations, respectively,

$$H_e = \Sigma_{\alpha n} \varepsilon_{\alpha n} a_{\alpha n}^\dagger a_{\alpha n} = \Sigma_{\alpha k} \varepsilon_\alpha(k) a_{\alpha k}^\dagger a_{\alpha k} \quad \text{(electron part)} \quad (2)$$

$$H_{ph} = \Sigma_n h\nu_\gamma(\mathbf{q}) b^\dagger_{\gamma \mathbf{q}} b_{\gamma \mathbf{q}} \quad \text{(phonon part)} \quad (3)$$

$$H_{int} = \Sigma_n G_n Q(\mathbf{n})(a_{\alpha n}^\dagger a_{\beta n} + a_{\beta n}^\dagger a_{\alpha n}) = \Sigma_{\mathbf{q}\gamma k} g_{\mathbf{q}\gamma k}(b_{\mathbf{q}\gamma} + b_{-\mathbf{q}\gamma}^\dagger)(a_{\alpha k}^\dagger a_{\beta k-q} + a_{\beta k}^\dagger a_{\alpha k-q}) \text{(coupling)} (4)$$

with

$$Q(\mathbf{n}) = [h/4\pi\nu_\gamma(\mathbf{q})]^{1/2} (b_{\mathbf{q}\gamma} + b_{-\mathbf{q}\gamma}^\dagger) \quad (5)$$

standing for the lattice (phonon) coordinate. Here $\alpha$ and $\beta$ are band labels, $\mathbf{q}$ is the phonon momentum, $\mathbf{k}$ is the electron momentum, $\gamma$ is the phonon band. The extension to boson (phonon) and fermion (electron) identification is made straightforwardly.

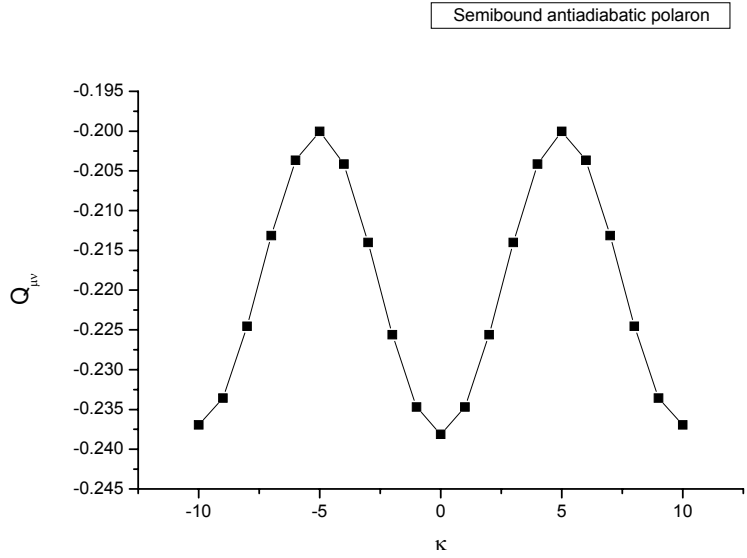

Figure 3

"Phonon coordinate" of a semibound antiadiabatic 1D vibronic polaron calculated by means of Merrifield's Variational Ansatz. See reference for a set of parameters used for the calculation. (Courtesy of reference [2]).

A closer inspection of the coupling term (interaction Hamiltonian) reveals that products of creation and annihilation operators of the form $b_{-q\gamma}^{\dagger} a_{\alpha k}^{\dagger} a_{\beta k-q}$, $b_{q\gamma} a_{\alpha k}^{\dagger} a_{\beta k-q}$, etc. signify the appearance of supersymmetric (boson-fermion) pairs $(b_{-q\gamma}^{\dagger}, a_{\alpha k}^{\dagger})$ or $(b_{q\gamma}, a_{\beta k-q})$ in the extended JT Hamiltonian. In as much as Hamiltonian (1) leads to the breaking of (external) symmetry the word can be of a link between external (configurational) symmetry breaking and super (internal) symmetry breaking. Indeed, the operators that dropped out in the above susy pairs, created $(b_{-q\gamma}^{\dagger}, a_{\alpha k}^{\dagger})$ and annihilated $(b_{q\gamma}, a_{\beta k-q})$, are $a_{\beta k-q}$ and $a_{\alpha k}^{\dagger}$, respectively. They signify fermion changes in the alternative bands.

The q-momentum dependence of the phonon coordinate (5) for a semibound vibronic polaron in 1D is reproduced in Figure 3 [2]. The coordinate is seen to be extremal at $q = \pm\frac{1}{2}\pi$ which points to a lower symmetry on both sides of the original higher symmetry at $q = 0$. Now, in as much as the Hamiltonian (1) is composed of operators belonging to the $q = 0$ symmetry, the lowered symmetry at $q = \pm\frac{1}{2}\pi$ would imply changing the phonon operators as functions of q and γ. This is the simplest case of phonon occurrence in JT phenomena, though it may be found instructive. One way or the other, a higher-symmetry phonon is destroyed while a lower-symmetry one is created in its place. This phonon transfiguring is reflected in related changes of the susy fermion-boson pair and in this sense the original pair is said to be changed while the supersymmetry is broken because its internal feature is transfigured. We see that in JT phenomena external and internal symmetries go hand in hand.

One way to reduce the effect of a symmetry breaking is to lower the interwell barrier as in Figure 1 depicting a double-well potential appearing in a semiclassical approach to the matter. In electron-phonon systems this has been done through increasing the interlevel energy gap, if the electron-phonon coupling strength has had to be kept fixed. If not, both gap energy and the coupling strength could be manipulated in order to control the barrier. For a low interwell barrier, the system easily performs flip-flop transitions from left to right which imitate or tend to restore on the average the broken configurational symmetry.

## 5. Conclusion

We have suggested that the most essential fingerprint of an off-centered electron-phonon system is the helical propagation waveform, else the appearance of an orbital angular momentum of the coupled system. We have also extended the conjecture to cover possibly any fermion-boson coupled system. Conceivable examples for such systems are provided by nucleon-$\pi$ meson coupling and by neutrino-meson coupling as well.

Of particular interest is the helical light propagation, else the orbital angular momentum of light beams, as observed for certain laser beams. Most important of them is the Laguerre-Gaussian mode and we suggest that this mode may provide the conditions for fermion (electron)-photon mixing in the active medium which is essential in our theory for the appearance of the helical form of light propagation.

Another essential appearance of the extended fermion-boson coupling is its possible relation to supersymmetry. In as much as the extended JT coupling may be expected to break the external symmetry of the system, it is accompanied by associated changes of its internal symmetry too. This makes the extended JT effects a strong candidate for the expected sypersymmetry breaking, presumed essential for understanding the scale of the fundamental interactions in particle physics.

The above extended JT effect story may sound as a fairy-tale, though it recognizes established facts that most of the longstanding problems in physics may allow interdisciplinary solutions. In any event, it would be too surprising if a solid state effect with so general a significance as the vibronic phenomenon could not account for some of the puzzling appearances in hep, or high energy physics.